\documentclass[referee]{aa} 
\usepackage[normalem]{ulem}
\usepackage{booktabs,caption}
\usepackage[flushleft]{threeparttable}
\usepackage{graphicx}
\usepackage{graphicx}
\usepackage{txfonts}
\begin{document}

   \title{X Persei: A study on the origin of its high-energy emission}
 \author{J. Rodi
          \inst{1}
          \and
          L. Natalucci
          \inst{1}
          \and
          M. Fiocchi
          \inst{1}
          }

   \institute{INAF - Istituto di Astrofisica e Planetologia Spaziali,
              via Fosso del Cavaliere 100; 00133 Roma, Italy\\
              \email{james.rodi@inaf.it}
             }

   \date{Received XXX; accepted XXX}

 
  \abstract
   {}
   {The origin of the hard X-ray emission in the Be/X-ray binary system X Persei has long been debated as its atypical ‘two-hump’ spectrum can be modelled in multiple ways.  The main debate focuses on the the high-energy hump, which is fit as either a cyclotron resonance scatter frequency (CRSF) or inverse Comptonization due to bulk Comptonization.}
   {Using \textrm{INTEGRAL}/JEM-X and ISGRI data, we studied the temporal and spectral variability in the \(3-250\) keV energy range during observations over \( \sim 15\) years.  A \textrm{NuSTAR} observation was also included in a joint spectral fit with the \textrm{INTEGRAL} spectrum.}
   {We find that the joint spectrum can be described well by a low-energy component due to thermal Comptonization and a high-energy component due to bulk Comptonization, a CRSF, or a cyclotron emission line.  The three models begin to diverge above \(\sim 120\) keV, where statistics are low.}
   {We compare our results with observations of other Be/X-ray binaries that show similar ‘two-hump’ spectra while in a low-luminosity state.  As the sources are in a low accretion state, the bulk Comptonization process is likely inefficient, and thus not an explanation for the high-energy component.  The broad CRSF (\(27 \pm 2\) keV) in X Persei suggests that the high-energy emission is not due to a CRSF. Thus, the high-energy component is potentially due to cyclotron emission, though other scenarios are not definitively excluded.}


   \maketitle
%

\section{Introduction}
The X-ray source 4U 0352\(+\)309 was first detected by the \textit{Uhuru} satellite \citep{giacconi1972} and was later found to be the companion of the Be star X Persei (X Per) \citep{braes1972,vandenbergh1972}.  The X-ray source has a pulse period of \( \sim 837 \) s \citep{white1976,haberl1998}, indicating that the compact object is a neutron star \citep{delgado2001}.  The system has an orbital period of approximately 250 days and has an inclination of \( \sim 26^{\circ}-33^{\circ}\) \citep{delgado2001}.  Distances to the system range from \(0.7 - 0.95\) kpc \citep{lyubimkov1997,roche1997,telting1998}.  \textit{Gaia} Data Release 3 reports a distance of \(0.69\) kpc.

Broadband observations at X-rays have predominately focused on understanding the emission process(es) responsible for the ‘two-hump’ spectrum.  Using \textit{Beppo}\textrm{SAX}, \cite{disalvo1998} described the spectrum as a thermal radiation component and a non-thermal partially Comptonization cyclotron line or high-temperature thermal bremsstrahlung.  \cite{coburn2001} analyses with \textit{RXTE} fit the spectrum as a blackbody spectrum with a power law modified by a cyclotron resonance scattering frequency (CRSF). \textrm{INTErnational Gamma-Ray Astrophysics Laboratory} (\textrm{INTEGRAL}) results from \cite{lutovinov2012} described the spectrum with a high-energy cut-off power law with a CRSF.  The thermal component is below the energy threshold, and thus was not included in the analysis.  

In contrast, a different \textrm{INTEGRA}L analysis reported that the spectrum can be fit as a thermal and bulk Comptonization model \citep{doroshenko2012}.  \textit{Suzaku} results provided intensity-dependent differences when fitting the spectra to the physically motivated \texttt{compag} model \citep{farinelli2012}.  Using \textit{Suzaku} data, \cite{maitra2017} found that the average and lower intensity spectra can be described well by just the \texttt{compag} model, but higher-flux spectra require a CRSF component.  

Though the origin of the spectral shape remains debated, similar two-hump spectra have been reported in other sources (1A 0535\(+\)262 \cite{tsygankov2019b}, GRO J1008\(-\)57 \cite{lutovinov2021}, and GX 304\(-\)1 \cite{tsygankov2019a}) while in a low-luminosity state (\( \sim 10^{34-35}\) erg/s).  \cite{sokolova2021} presented an alternative model to explain the GX 304\(-\)1 spectrum as a low-energy hump due to a thermal component and a high-energy component due to resonant Comptonization.  Thus, the process(es) at work in those sources is/are likely the same as in X Per, which has been observed at the same luminosity level \citep{white1976,lyubimkov1997,disalvo1998,coburn2001,palombara2007,lutovinov2012,maitra2017}.

In this work, we investigate \textrm{INTEGRAL} observations over the course of the mission (\( \sim 15\) years) to search for spectral variability and differentiate between previously proposed physical models for Be/X-ray binaries in a low-luminosity state.

\section{Observations and data reduction}

\textrm{INTEGRAL} was launched into an elliptical orbit in October 2002 from Baikonur, Kazakhstan \citep{jensen2003}.  The instruments on board \textrm{INTEGRAL} have observed X Per numerous times during the course of the mission, providing a long baseline of observations.  In this work, we analyzed \textrm{INTEGRAL} observations using data from the \textrm{Joint European Monitor for X-rays} (\textrm{JEM-X}) \citep{lund2003} and the \textrm{INTEGRAL on-board Imager} (\textrm{IBIS}) \citep{ubertini2003}.  The data span \textrm{INTEGRAL} revolutions \(170-2122\) (MJD \(53069-58702\)).  There are 1327 \textrm{IBIS} observations (science windows) during this period for a total exposure time of 2250 ks and 364 science windows (scw) with \textrm{JEM-X} for a total exposure time of 830 ks.  \textrm{JEM-X} has fewer observations due to its smaller field of view (FoV) compared to IBIS and the fact that many observations including X Per do not have the source as the main target.  


The \textrm{JEM-X} and \textrm{IBIS}/\textrm{ISGRI} (\textrm{INTEGRAL Soft Gamma-Ray Imager}) analyses were performed using the standard Offline Science Analysis (OSA) tools developed for \textrm{INTEGRAL} data.\footnote{https://www.isdc.unige.ch/integral/analysis\#Software} Following the JEM-X Analysis User Manual, JEM-X spectral analysis was performed using \textit{jemx\_science\_analysis} (v11.1) with 49 channels in the \(3-20\) keV energy range.  For ISGRI,  spectral analysis was performed using the routine \textit{ibis\_science\_analysis} (v11.2), as is documented in the IBIS Analysis User Manual.  The spectra were extracted in 28 channels in the \(30-250\) keV energy range.  Average spectra for each instrument were generated using the OSA tool \textit{spe\_pick}.  A 1\% systematic error was added, as is suggested for standard analyses.

For \textrm{NuSTAR}, the data are from observation 30401033002, corresponding to 2019-01-01 10:26:09 to 2019-01-02 20:11:09 UTC.  As is outlined in the \textrm{NuSTAR} Data Analysis Software Guide,\footnote{https://heasarc.gsfc.nasa.gov/docs/nustar/analysis/nustar\_swguide.pdf} data reduction for FPMA and FPMB was performed with the \textit{nuproducts} routine using \textit{nustards\_01Apr\_v1.9.2} and CALDB version 2022-01-29.  The total exposure time is \(38.4\) ks.  For image analysis, the source extraction used a 116 arcsec circular region and a 142 arcsec background circular region. Spectra were generated in the \(3-70\) keV energy range and rebinned to 800 bins (FPMA) and 782 bins (FPMB) to maximize the signal-to-noise ratio at high energies.  

\section{Results}

\subsection{Temporal results}

\begin{figure*}
    \sidecaption
    \includegraphics[scale=0.70, angle=180,trim = 10mm 10mm 40mm 0mm, clip]{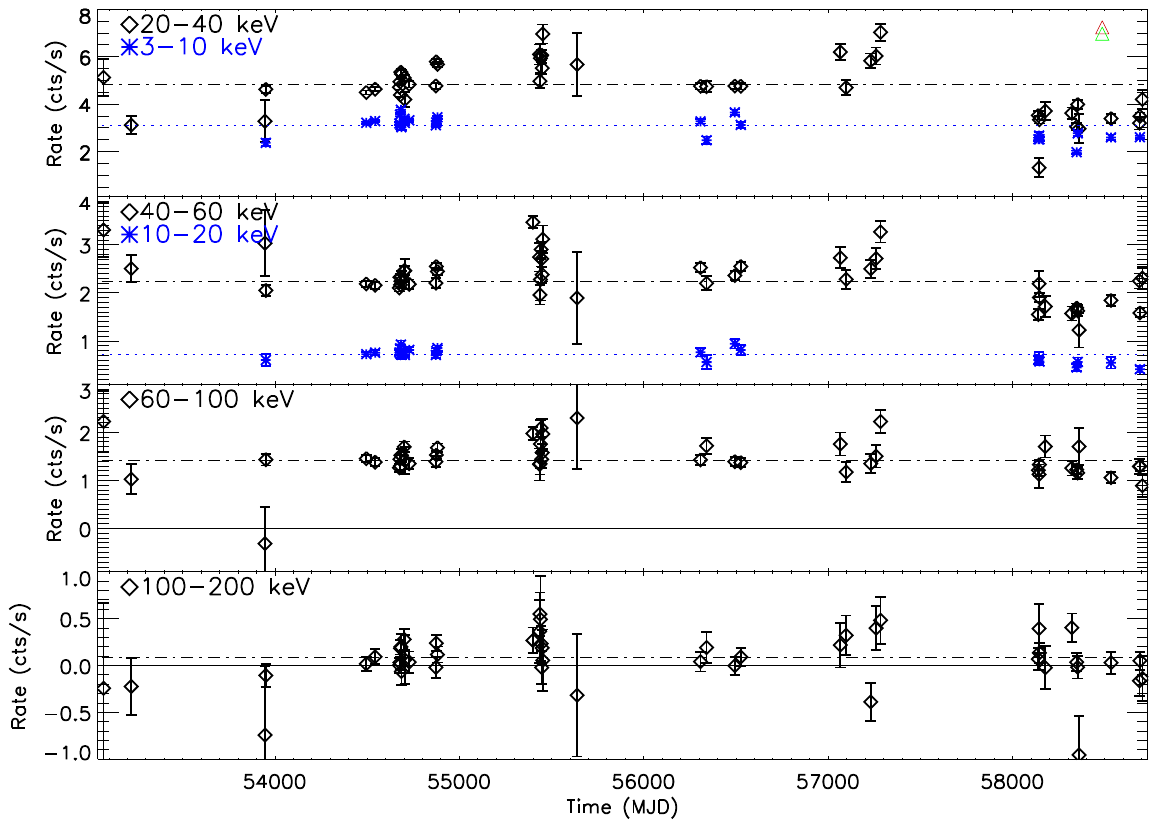}
    \caption{X Per long-term light curve with \textrm{JEM-X} in the \(3-10\) keV and \( 10-20 \) keV energy bands as blue stars, and \textrm{ISGRI} in the \(20-40\) keV, \(40-60\) keV, \(60-100\) keV, and \(100-200\) keV energy bands.  The average \textrm{JEM-X} count rate is shown as a dotted blue line.  The average \textrm{ISGRI} count rate is shown as a dash-dotted black line.  The value of 0 cts/s is shown as a solid black line.  The \textit{NuSTAR} \(3-79\) keV count rates are shown as a red and green triangles and have been normalized by a factor of 0.5.}
    \label{fig:xper_lc}
\end{figure*}

The long-term light curves of \textrm{JEM-X} and \textrm{ISGRI} are shown in Figure~\ref{fig:xper_lc}.  The points are on an \textrm{INTEGRAL} revolution timescale (\(\sim 2-3\) days).  This timescale maximizes the signal-to-noise ratio when searching for long-term spectral variability below in Sec~3.2.1.  The \textrm{JEM-X} points are shown as blue stars.  The \textrm{ISGRI} points are show as black diamonds.  The \textit{NuSTAR} \(3-79\) keV count rates for FPMA and FPMB are \(14.50 \pm 0.02\) and \(13.96 \pm 0.02\) ct/s, respectively, and are plotted as red (FPMA) and green (FPMB) triangles.  The errors are \(\pm 1\sigma\), as are all subsequent errors.  The \textit{NuSTAR} count rates have been normalized by a factor of 0.5.  The average \textrm{JEM-X} count rate is shown as a dotted blue line.  The average \textrm{ISGRI} count rate is shown as a dash-dotted black line.  The value of 0 cts/s is shown as a solid black line.

X Per shows temporal variability over the course of the mission and was significantly detected up to 100 keV by \textrm{INTEGRAL}.  The average count rates (and significances) in \textrm{JEM-X} are \(3.11 \pm 0.02\) ct/s (\(202.8 \sigma\)) and \(0.73 \pm 0.01\) ct/s (\( 52.3 \sigma \)) in the \(3-10\) keV and \( 10-20 \) keV energy bands, respectively.  The \textrm{ISGRI} count rates (and significances) are \(4.84 \pm 0.02\) ct/s (\(200.0 \sigma\)), \(2.23 \pm 0.02\) ct/s (\(128.0 \sigma\)), \(1.42 \pm 0.02\) ct/s (\(74.4 \sigma\)) in the \(20-40\) keV, \(40-60\) keV, and \(60-100\) keV energy bands, respectively.  The source is marginally detected at \(4.5 \sigma\) in the \(100-200\) keV energy band with \(0.09 \pm 0.02\) ct/s.

\subsection{Spectral results}

\subsubsection{Spectra variability}

\begin{figure*}[!h]
    \sidecaption
	\includegraphics[scale=0.8, angle=0,trim = 25mm 20mm 10mm 80mm, clip]{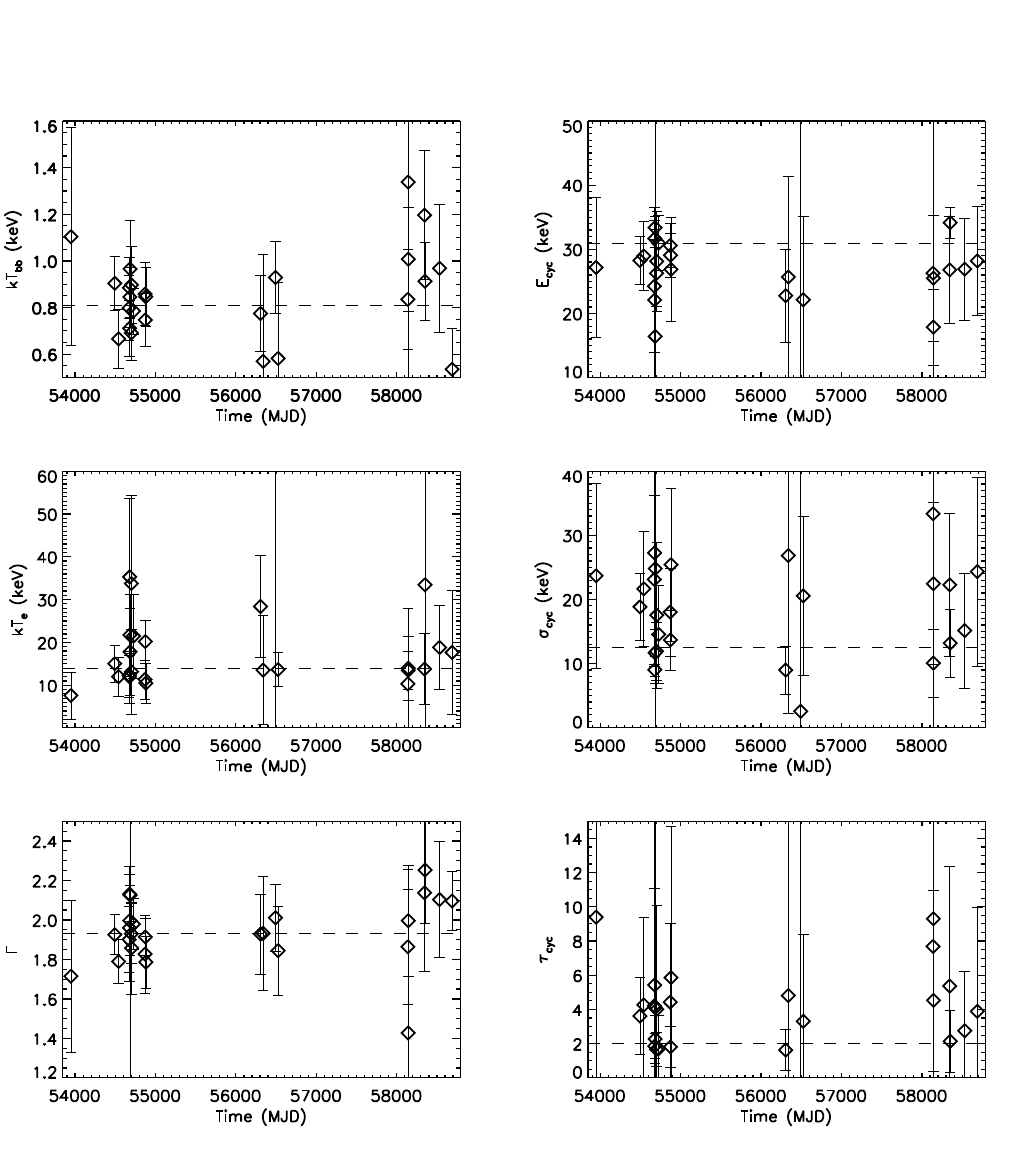}
    \caption{Evolution of X Per best-fit parameters to a \texttt{nthcomp*gabs} model on a revolution timescale.  The weighted average value is overplotted as a dashed line in each panel.}
    \label{fig:param_var}
\end{figure*}


To search for spectral variability on a revolution timescale (\(\sim 2 - 3\) d), we fitted joint JEM-X/ISGRI spectra using XSPEC version 12.9.0.  Previous works on X Per modelled the spectrum as a power law with a high-energy cut-off (\texttt{powerlaw*highecut}) modified by a CRSF (\texttt{gabs}) \citep{coburn2001,lutovinov2012}.  Thus, we fitted the spectra with an \texttt{nthcomp*gabs} model as it provides a more physically motivated continuum spectrum with the same number of free parameters.  In Sec~\ref{sec:avg_spec} below, we evaluate the merits of different spectral models.  Figure~\ref{fig:param_var} shows the revolution timescale parameters with the continuum parameters in the first column and the \texttt{gabs} parameters in the second column.  The average values for each parameter are plotted as a dashed line in each panel.  The disk blackbody (\( kT_{bb}\)), electron temperature (\(kT_e\)), and the power law index  (\( \Gamma \)) appear constant with average values of \(0.81 \pm 0.03 \) keV, \( 14 \pm 1\) keV, and \(1.93 \pm 0.03\), respectively.  The \texttt{gabs} parameters show a similar behavior with no clear long-term variability in the cyclotron energy, line width, or optical depth, respectively.  The average values are \(E_{cyc} = 30.9 \pm 0.8\) keV, \(\sigma_{cyc} = 12 \pm 1\) keV, and \(\tau_{cyc} = 2.0 \pm 0.4\)).  However, the errors on the parameters of an individual revolution timescale are often large due to relatively short exposure times. 
 
 
 For the JEM-X spectra, the exposure time varies from \( \sim 7.7 - 66\) ks.  For the ISGRI spectra, the exposure time varies from \( \sim 10 - 125\) ks.

\begin{figure}
	\includegraphics[scale=0.8, angle=180,trim = 35mm 70mm 65mm 20mm, clip]{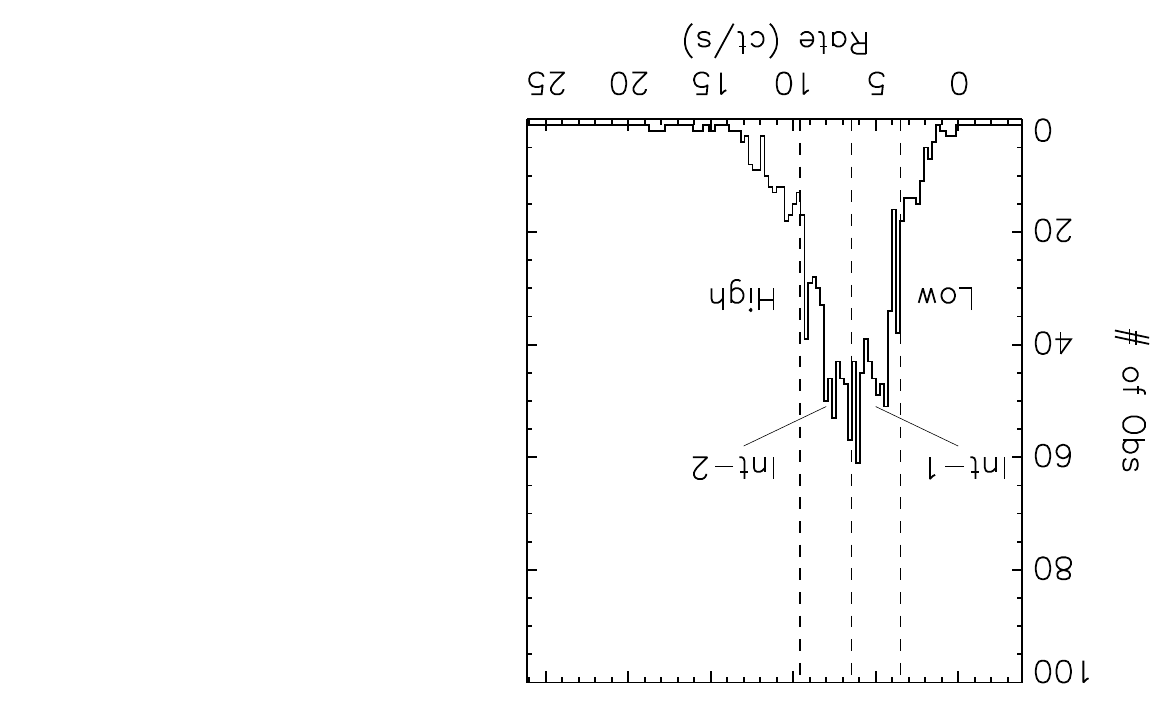}
    \caption{Histogram of the \(20-80\) keV count rate with dashed lines denoting low, intermediate-1, intermediate-2, and high count-rate levels.}
    \label{fig:flux_hist}
\end{figure}
\begin{figure*}
	\includegraphics[scale=0.8, angle=0,trim = 20mm 25mm 5mm 70mm, clip]{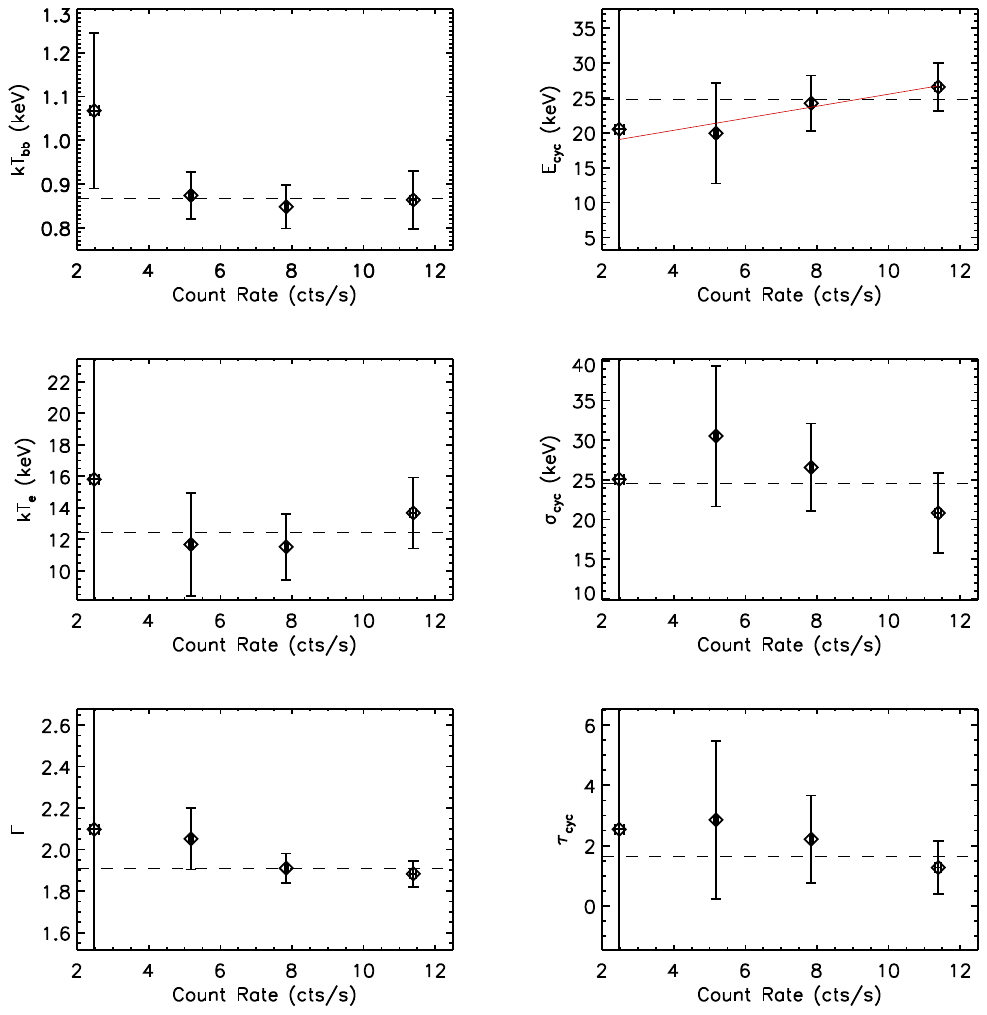}
    \caption{Evolution of X Per best-fit parameters to a \texttt{nthcomp*gabs} model for different flux levels.  The average values are overplotted as a dashed line in each panel.  The red line in the \(E_{cyc}\) panel is a fit to the cyclotron energy vs count rate.}
    \label{fig:flux_level}
\end{figure*}

To reduce the fit parameter errors, we built a count-rate histogram using the \(20-80\) keV count rate (Figure~\ref{fig:flux_hist}), as \cite{maitra2017} and grouped the data into four count-rate levels: low, intermediate-1, intermediate-2, and high.  As an initial test, we fitted the spectra to the \texttt{nthcomp} model.  In each case, the spectrum was poorly described by the model with \(\chi^2 / \nu \) values of \( 137.90 / 67 = 2.06\), \(390.13/67 = 5.82\), \(460.21/67 = 6.87\), and \(242.67/67 = 3.62\) for the low, intermediate-1, intermediate-2, and high count-rate levels, respectively.  Subsequently, we fitted the spectra using the \texttt{nthcomp*gabs} model and found that each spectrum is described well by the data (See Table~\ref{table:1}).  We also tested the need for an additional component (that is modelled with \texttt{gabs}) using \texttt{simftest} in Xspec for each count-rate level.  Using 1,000 iterations, the fit did not improve in any of the realizations.  Thus, an additional component is required at a level of \( \gtrsim 3.3 \sigma\) (1/1000).  The best-fit parameters for the \texttt{nthcomp*gabs} model are listed in Table~\ref{table:1}.  The best-fit parameters versus the mean count rate within a bin are plotted in Figure~\ref{fig:flux_level}.  The average fit parameters are plotted as a dashed line in each panel.  

The continuum parameters \textbf{(\( kT_{bb} \), \( kT_{e}\), \( \Gamma \))} do not show any significant dependence on flux, though the errors during the low flux period are large.  A positive correlation between flux and cyclotron absorption energy has been reported in a number of sources (e.g. \cite{staubert2007,klochkov2012,decesar2013,furst2014,sartore2015}); thus, the presence of such a correlation would support an interpretation of a CRSF in X Per.  However, the cyclotron absorption parameters (E\(_{cyc}\), \(\sigma_{cyc}\), and \(\tau_{cyc}\)) also do not exhibit significant flux evolution.  The E\(_{cyc}\) values are consistent with the average value.  We fitted the data to a linear model.  The best-fit slope was \textbf{\( 0.9 \pm 1.0\)} (shown as a red line in the top panel of the second column of Figure~\ref{fig:flux_level}), consistent with a slope of 0.  Therefore, we consider the X Per spectrum to be constant during \textit{INTEGRAL} observations in this work. 


\begin{table*}
\begin{center}
\caption{Count-rate level fit parameters with \texttt{Tbabs*(nthcomp*gabs)}}
 \hspace{0mm}
\scalebox{0.995}{
\hspace{-6mm}
\begin{tabular}{cccccc}
\hline                   \\
                       & Low                   & Intermediate-1         & Intermediate-2          & High \\           \hline\\
\(kT_{bb}\)            & \(1.1 \pm 0.2\) keV   &\(0.87 \pm 0.05\) keV   & \(0.85 \pm 0.05\) keV   & \(0.9 \pm 0.7\) keV     \\
\(kT_{e}\)             & \(16  \pm 8\) keV     &\(12 \pm 3\) keV        & \(12 \pm 2\)  keV       & \(14 \pm 2\) keV         \\ 
\(\Gamma\)             & \(2.1 \pm 0.6\)       &\(2.1 \pm 0.1\)         & \(1.9 \pm 0.7\)        & \(1.88 \pm 0.06\)        \\ 
\(E_{cyc}\)            & \(20 \pm 17\) keV     &\(20 \pm 7\) keV        & \(24 \pm 4\) keV        & \(27 \pm 3\) keV         \\
\(\sigma_{cyc}\)       & \(25 \pm 15\) keV     &\(31 \pm 9\) keV        & \(27 \pm 6\) keV        & \(21 \pm 5\) keV         \\ 
\(\tau_{cyc}\)         & \(3 \pm 4\)           &\(3 \pm 3\)             & \(2 \pm 1\)             & \(1.3 \pm 0.9\)          \\ 
\hline \\
red. \( \chi^2 / \nu \)& \(0.90/64\)           &\(0.83/64\)             & \(0.59/64\)             & \(0.77/64\)              \\ 

\hline
\end{tabular}
}

\label{table:1}

\end{center}
 \begin{tablenotes}
     \item [-] -\texttt{nthcomp} mode: \(kT_{bb}\) is the seed photon temperature, \(kT_e\) is the electron temperature, and \(\Gamma\) is the power law index.
    \item [-] -\texttt{gabs} model: \(E_{cyc}\), \(\sigma_{cyc}\), and \(\tau_{cyc}\) are the energy, width, and optical depth of the CRSF line, respectively.
     \end{tablenotes}
\end{table*}

\label{table:1}


There is a 10 keV gap between JEM-X and ISGRI (\(20-30\) keV).  To bridge this energy range, spectra from a \textit{NuSTAR} observation were included.  As a test of agreement between the instruments, we fitted the average \textit{INTEGRAL} spectra with the \textit{NuSTAR} spectra and found a red. \( \chi^2/d.o.f. = 0.99/1647\).  Thus, the four spectra are in good agreement.  The instrument cross-normalizations were \(C_{JEM-X} = 1.05 \pm 0.01\), \(C_{FPMA} = 0.965 \pm 0.008\), and \(C_{FPMB} = 0.944 \pm 0.008\).  The ISGRI cross-normalization was fixed to 1.  The fit results are discussed in the subsequent section.  


\subsubsection{Average spectrum}~\label{sec:avg_spec}

Due to the uncertain nature of the hard X-ray emission, the source has been studied by multiple missions over the past decades.  Results can generally be grouped into models using two additive components \citep{disalvo1998,doroshenko2012} and models with a CRSF \citep{coburn2001,lutovinov2012,maitra2017}.  We began with spectral fits to the JEM-X/ISGRI/\textit{NuSTAR} data using phenomenological models.  A fit to an absorbed power law model with a cross-instrument normalization \texttt{constant*Tbabs*powerlaw} found \( \Gamma = 2.095 \pm 0.001\) (red. \(\chi^2 /d.o.f =\) \(14.11 / 1650)\).  nH was fixed to \(0.15 \times 10^{22} \textrm{ cm}^{-2}\) based on \cite{coburn2001}.  The residuals are complex, with the model over-predicting the observed flux \(~< 5\) keV, between \( ~10 - 40\) keV, and \(~> 90\) keV, suggesting the presence of a cut-off power law continuum spectrum or possibly a more complex continuum.  An absorbed cut-off power law model finds best-fit parameters of \( \Gamma = 1.879 \pm 0.002\), a cut-off energy with \( E_C = 6.6 \pm 0.1\) keV, and a fold energy of \(E_{fold} = 36.8 \pm 0.5\) keV (red. \(\chi^2 /d.o.f =\) \(8.99 / 1648)\).  The residuals show that the model again over-predicts the observed flux at energies similar to the power law model.  

Following \cite{disalvo1998}, we fitted the spectrum using a phenomenological two-component model with two power laws with high-energy cut-off components with low-energy and high-energy cut-offs.  The \texttt{constant*Tbabs*(powerlaw*highecut+powerlaw*highecut)} fit finds similar parameters, which are listed in Table~\ref{table:paras1}, and acceptably fit the data (red. \(\chi^2 /d.o.f =\) \(1.11 / 1644)\).  The cross-normalization parameters relative to the ISGRI flux are \(C_{JEM-X} = 1.05 \pm 0.01\), \(C_{FPMA} = 0.967 \pm 0.008\), and \(C_{FPMB} = 0.946 \pm 0.008\).  Fits with subsequent models resulted in similar cross-normalization values.  The JEM-X and ISGRI cross-normalizations are in good agreement.  The \textit{NuSTAR} cross-normalization values are close to 1.  However, the spectra come from a single observation compared to the long-term average for ISGRI; thus, values different from 1 are unsurprising.


A model with two continuum components is able to adequately explain the spectrum.  However, other authors \citep{coburn2001,lutovinov2012,maitra2017} have proposed that the dip feature is due to a CRSF instead of two different hard X-ray emission components.  Thus, we fitted the spectrum to a \texttt{constant*Tbabs(powerlaw*highecut*gabs)} model.  The best-fit parameters are similar to previous results and are given in Table~\ref{table:paras1}; again, they describe the data well (red. \(\chi^2 /d.o.f =\) \(1.02 / 1645)\).

\begin{table*}
\begin{center}
\caption{Phenomenological continuum spectrum parameters}
 \hspace{0mm}
\scalebox{0.995}{
\hspace{-6mm}
\begin{tabular}{cccccc}
\hline \\
\multicolumn{2}{c}{\texttt{Tbabs*(powerlaw*highecut+powerlaw*highecut)}} & \multicolumn{1}{c}{}  & \multicolumn{2}{c}{\texttt{Tbabs*(powerlaw*highecut*gabs)}} &  \\
\hline \\
\(\Gamma_1\)            & \(0.80 \pm 0.02\)     &\(\) & \(\Gamma\)       & \(2.5 \pm 0.6\)        \\
\(E_{cut_1}\)           & \(3.30 \pm 0.05\) keV &\(\) & \(E_{cut}\)      & \(5.7 \pm 0.5\) keV    \\ 
\(E_{fold_1}\)          & \(5.0 \pm 0.1\) keV   &\(\) & \(E_{fold}\)     & \(17.8 \pm 0.6\) keV   \\ 
\(\Gamma_2\)            & \(1.40 \pm 0.02\)     &\(\) & \(E_{cyc}\)      & \(17.8 \pm 0.6\) keV   \\
\(E_{cut_2}\)           &\(70 \pm 1\) keV       &\(\) & \(\sigma_{cyc}\) & \(29.2 \pm 0.8\) keV   \\ 
\(E_{fold_2}\)          & \(23 \pm 2\) keV      &\(\) & \(\tau_{cyc}\)   & \(3.6 \pm 0.3\)        \\ 
\hline \\
red. \( \chi^2 / \nu \) & \(1.11/1644\)         &\(\) & \(\)             & \( 1.02/1645\)         \\ 

\hline
\end{tabular}
}

\label{table:paras1}

\end{center}
 \begin{tablenotes}
     \item [-] -\texttt{powerlaw} mode: \(\Gamma\) is the power law index
    \item [-] -\texttt{highecut} model: \(E_{cut}\) is the cut-off energy, \(E_{fold}\) is the e-fold energy
    \item [-] -\texttt{gabs} model: \(E_{cyc}\), \(\sigma_{cyc}\), and \(\tau_{cyc}\) are the energy, width, and optical depth of the CRSF line, respectively.
     \end{tablenotes}
\end{table*}


\begin{figure*}
	\includegraphics[scale=0.815, angle=0,trim = 10mm 55mm 20mm 40mm, clip]{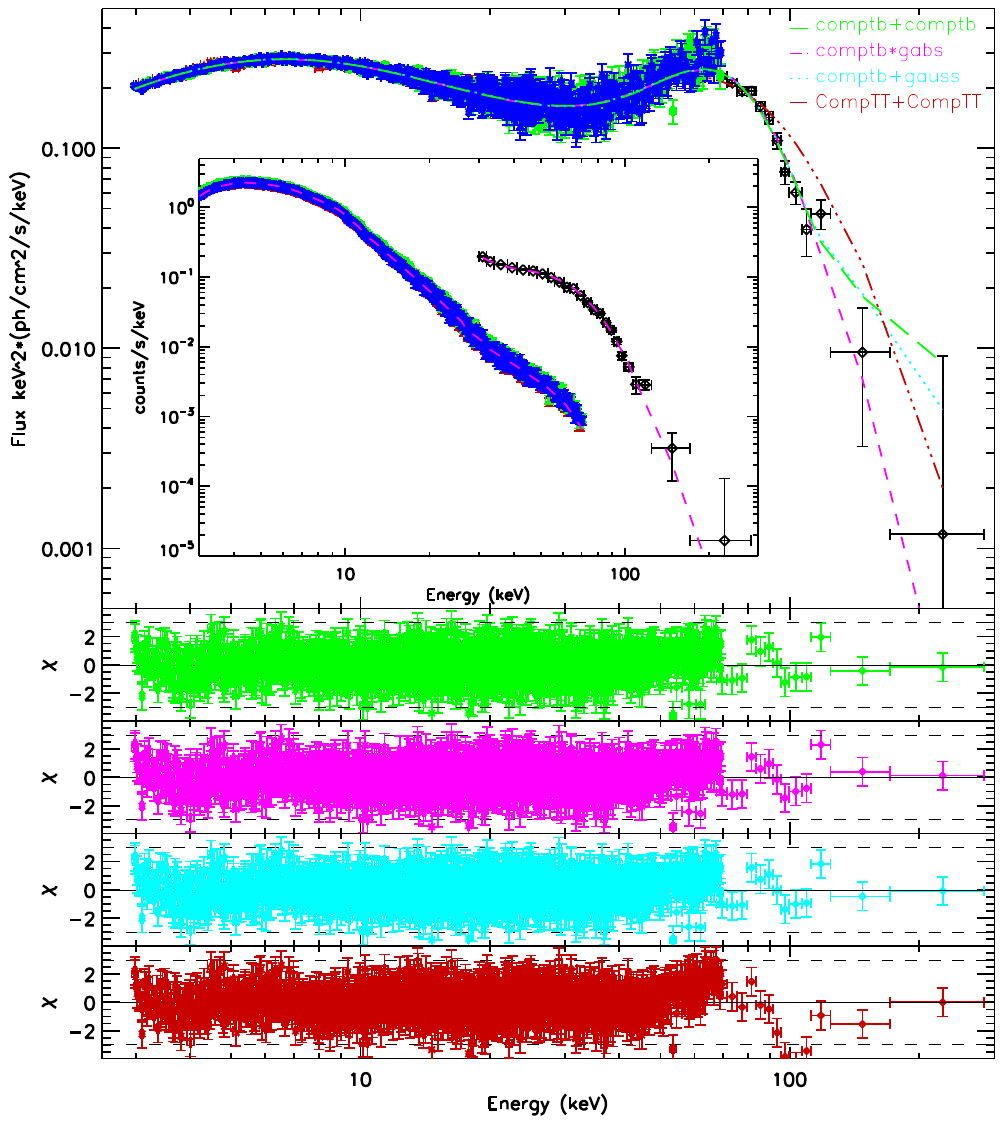}
    \caption{X Per average spectrum with \textrm{ISGRI} (black diamonds), \textrm{JEM-X} (red triangles), \textit{NuSTAR}/FPMA (blue stars), and FMPB (green squares).  The best-fit models are overplotted with \texttt{comptb+comptb} (long green dashes), \texttt{comptb*gabs} (short magenta dashes), \texttt{comptb+gauss} (cyan dots), and \texttt{CompTT+CompTT} (red dots and dashes).  The folded spectrum with the \texttt{comptb*gabs} model is shown as an inset.  Plots of the residuals for the \texttt{comptb+comptb}, \texttt{comptb*gabs}, \texttt{comptb+gauss}, and \texttt{CompTT+CompTT} fits, respectively, are plotted in panels below the spectrum.  Dashed lines are plotted at \(\pm 3 \sigma\).}
    \label{fig:avg_spec}
\end{figure*}

Next, we tried physically motivated models to explain the origin of the two ‘humps.’  Following the analysis of \cite{doroshenko2012}, we also fitted the data to a \texttt{CompTT+CompTT} model, which attributes the ‘humps’ to thermal Comptonization processes.  In this model, the free parameters are the photon seed temperature (\(kT_0\)), the electron temperature (\(kT_e\)), and the optical depth (\(\tau\)).  A two \texttt{CompTT} model provides a statistically acceptable fit to the model (red. \(\chi^2 /d.o.f =\) \(1.06 / 1646)\) but is qualitatively worse, especially at high energies (see Figure~\ref{fig:avg_spec} and discussion below.) The best-fit parameters are given in Table~\ref{table:paras2}, as are the parameters for subsequent fits.  

\cite{becker2005a,becker2005b} proposed that accreting X-ray pulsar spectra are due to a combination of thermal and bulk Comptonization.  Thus, we fitted the data with a \texttt{comptb+comptb} model \cite{farinelli2008}.  In the \texttt{comptb} model, the free parameters are the seed temperature (\(kT_s\)), the seed photon spectral index (\(\Gamma\)), the Comptonization spectrum energy index (\(\alpha\)), the efficiency of the bulk Comptonization to the thermal Comptonization (\(\delta\)), the electron temperature (\(kT_e\)), and the illumination factor (\(A\)), which is the related to the fractional amount of seed photon radiation directly observed.  An initial fit with all the parameters free is able to describe the data well (red. \(\chi^2/d.o.f = 1.04/1638\)), but none of the parameters  are constrained.  Following the analysis of accreting neutron stars in \cite{farinelli2008}, we fixed \( \delta\) to 0 and \(A\) to 8 for a low-energy component due \textcolor{red}{to} thermal emission.  In this scenario, the low-energy \texttt{comptb} model simplifies to the \texttt{CompTT} model \cite{farinelli2008}.  The model describes the data well (red. \(\chi^2 /d.o.f =\) \(0.98 / 1644)\) with significant bulk Comptonization found (\(\delta = 13 \pm 2\)), though the several of the parameters are unconstrained or poorly constrained.  This model provides a better description of the data compared to the \texttt{CompTT+CompTT} model with \( \Delta \chi^2 = 126.51\) with two fewer degrees of freedom.

Subsequently, we tested a \texttt{comptb*gabs} model to possibly explain the spectrum with CRSF.  As before, delta was fixed to 0, and the log of the illumination was fixed to 8.  The model also fits the data well (red. \(\chi^2 /d.o.f =\) \(0.99 / 1646)\).  The CRSF model (\( \chi^2/\nu = 1631.66/1646 \)) provides a better fit than the two \texttt{CompTT} model with \( \Delta \chi^2 = 112.65 \) with the same number of free parameters and a comparable \( \chi^2\) to \texttt{comptb+comptb} (\( \Delta \chi^2 = -13.86\)), though with two more free parameters.

\cite{nelson1993} report that for low-luminosity accreting X-ray pulsars such as X Per, a broad cyclotron emission line may be present below the cyclotron energy.  Thus, we fitted the data to a \texttt{comptb+gauss} model assuming that the second ‘hump’ is a cyclotron emission line instead of a cyclotron absorption line.  As in the \texttt{comptb*gabs} model, delta was fixed to 0 and the log of the illumination was fixed to 8.  Again, we find the data are described well by the model (red. \(\chi^2 /d.o.f =\) \(0.98 / 1644)\).  The cyclotron emission line model (\( \chi^2/\nu = 1625.63/1646 \)) provides a better fit than the two \texttt{CompTT} model (\( \Delta \chi^2 = 118.68\)) for two more free parameters and a comparable \( \chi^2\) to the two \texttt{comptb} and \texttt{comptb*gabs} models with \( \Delta \chi^2 \) values of \(-7.83\) and \(6.03\), respectively, for the same number of free parameters.

The average spectrum and best-fit models are shown in Figure~\ref{fig:avg_spec} with the \textrm{ISGRI} (black diamonds), \textrm{JEM-X} (red triangles), \textit{NuSTAR}/FPMA (blue stars), and FMPB (green squares) data.  The best-fit models are overplotted with \texttt{comptb+comptb} (long green dashes), \texttt{comptb*gabs} (short magenta dashes), \texttt{comptb+gauss} (cyan dots), and \texttt{CompTT+CompTT} (red dots and dashes).  The folded spectrum with the \texttt{comptb*gabs} model is shown as an inset.  The panels below the spectrum show the residuals to the \texttt{comptb+comptb}, \texttt{comptb*gabs}, \texttt{comptb+gauss}, and \texttt{CompTT+CompTT} fits, respectively.  Dashed lines are plotted at \(\pm 3 \sigma\).

\begin{table*}
\begin{center}
\caption{Physically motivated spectrum parameters  }
 \hspace{0mm}
\scalebox{0.995}{
\hspace{-6mm}
\begin{tabular}{ccccccccc}
\hline \\
\multicolumn{2}{c}{\texttt{Tbabs*(comptb+comptb)}} &  \multicolumn{1}{c}{}  & \multicolumn{2}{c}{\texttt{Tbabs*(comptt+comptt)}}   & \multicolumn{2}{c}{\texttt{Tbabs*(comptb*gabs)}} &  \multicolumn{2}{c}{\texttt{Tbabs*(comptb+gauss)}}  \\
\hline \\
\(kT_s\)                & \(2.0 \pm 0.2\) keV &\(\) &\(kT_{0_1}\)   &\(0.57 \pm 0.02\) keV   &\(kT_s\)         & \(1.86 \pm 0.01\)    & \(kT_s\)           & \(1.77 \pm 0.05\) keV       \\
\(\Gamma_{1}\)          & \(1.4 \pm 0.8\)     &\(\) & \(kT_{e_1}\)  &\(2.25 \pm 0.03\) keV   &\(\Gamma\)       & \(1.58 \pm 0.06\)    & \(\Gamma\)         & \(1.54 \pm 0.04\)     \\ 
\(\alpha_{1}\)          & \(1.8 \pm 0.1\)     &\(\) & \(\tau_1\)    &\(8.3 \pm 0.1\)         &\(\alpha\)       & \(1.10 \pm 0.04\)    & \(\alpha\)         & \(1.55 \pm 0.02\)    \\ 
\(kT_{e_1}\)            & \(57 \pm 92\) keV   &\(\) & \(kT_{0_2}\)  &\(14.7 \pm 0.1\) keV    &\(kT_e\)         & \(10.8 \pm 1.0\) keV & \(kT_e\)           & \(28 \pm 6\) keV   \\ 
\(\Gamma_2\)            & \(1.2 \pm 24.4\)    &\(\) & \(kT_{e_2}\)  &\(3.06 \pm 0.03\) keV   &\(E_{cyc}\)      & \(23 \pm 1\) keV     & \(E_{gauss}\)      & \(51.4 \pm 0.8\) keV   \\
\(\alpha_{2}\)          & \(0.39 \pm 0.05\)   &\(\) & \(\tau_2\)    &\(20.0 \pm 0.5\)        &\(\sigma_{cyc}\) & \(27 \pm 2\) keV     & \(\sigma_{gauss}\) & \(21.4 \pm 0.9\) keV      \\ 
\(kT_{e_2}\)            & \(4.2 \pm 0.5\) keV &\(\) & \(\)  &\(\)                            &\(\tau_{cyc}\)   & \textbf{\(2.4 \pm 0.7\)}     & \(\)               & \(\)    \\ 
\(\delta\)              & \(13 \pm 2\)        &\(\) & \(\)  &\(\)                            &\(\) & & \(\)               & \(\)    \\ 
\hline \\
red. \( \chi^2 / \nu \) & \(0.98/1644\)       &\(\) & \(\)\(\) &\(1.06/1646\)                &                 &\(0.99/1646\) &             & \( 0.99/1646\)         \\ 

\hline
\end{tabular}
}
\label{table:paras2}

\end{center}
 \begin{tablenotes}
     \item [-] -\texttt{comptb} model: \(kT_s\) is the seed photon temperature, \(\ \Gamma\) is the seed photon spectral index, \( \alpha\) is the Comptonization spectrum energy index, \(kT_e\) is the electron temperature, and \(\delta\) is the efficiency of the bulk Comptonization to the thermal Comptonization.  
    \item [-] -\texttt{compTT} model: \(kT_{0}\) is the seed photon temperature, \(kT_e\) is the electron temperature, and \(\tau \) is the optical depth.    \(E_{gauss}\) and \(\sigma_{gauss}\) are the energy and width of the cyclotron emission line, respectively.  
    \item [-] -\texttt{gabs} model: \(E_{cyc}\), \(\sigma_{cyc}\), and \(\tau_{cyc}\) are the energy, width, and optical depth of the CRSF line, respectively.
    \item [-] -\texttt{gauss} model: \(E_{gauss}\) and \(\sigma_{gauss}\) are the energy and width of the cyclotron emission line, respectively.
     \end{tablenotes}
\end{table*}

\section{Discussion}

\subsection{Comparison of model results}

In previous works, many of the analyses reporting a CRSF used only phenomenological models with a cut-off power law model.  Because a cut-off power law model decreases differently from a Comptonization model \citep{petrucci2001}, we focused on physically motivated models to fit the average spectrum.  Our fit to a two \texttt{CompTT} model (\( \chi^2/\nu = 1744.31 / 1646\)) is able to describe the spectra well until \(\sim 95\) keV.  Above this energy, the model over-predicts the observed flux (See Figure~\ref{fig:avg_spec}.), suggesting that the \texttt{CompTT+CompTT} model is unable to describe the high-energy spectrum.

The \texttt{comptb+comptb} model provides a better description of the data.  This model found significant bulk Comptonization (\(\delta = 13 \pm 2\)), implying an optically thick environment and a high accretion rate \citep{tsygankov2019a}, which is unlikely to be present in the low-luminosity state \citep{mushtukov2015}.  The model is above the observed data for the last two data points, but the errors on those points are large. 

The CRSF model provides a better fit than the two \texttt{CompTT} model and comparable to the \texttt{comptb+comptb}.    However, the broad cyclotron line (\( \sigma = 27 \pm 2\) keV) is quite broad compared to typical cyclotron lines \citep{coburn2002,staubert2019}.  This could be an explained by an artificial broadening due to averaging over a variable \(E_{cyc}\).  As is shown in Figure~\ref{fig:flux_level}, there is no long-term variability of the cyclotron line position.  Additionally, phase analysis of the pulsar did not find significant variability in the line position \citep{coburn2001}, which means short-term variability is not the cause of the broad cyclotron line.  

\cite{coburn2001} attribute the broadened line to viewing the source off-angle at \(23-30^{\circ}\) \citep{delgado2001}.  They assume an electron temperature of \( \geq 60\) keV, using the best-fit parameter from a cut-off power law model, significantly lower than the \(kT_e = 10.9\) keV found in this work.  Using the equation from \cite{meszaros1985}, which accounts for broadening to thermalized electrons and Doppler broadening, with the values from our analysis, we find an expected width of roughly 3 keV, which is inconsistent with the 27 keV based on our spectral analysis.  We note that the \texttt{gabs} parameters depend on the assumed continuum model. 

Finally, the cyclotron emission line model provides a statistical fit comparable to the \texttt{comptb+comptb} and the CRSF models.  However, the line energy (\(\sim 51\) keV) is significantly higher than predicted (\(E_{gauss} \sim 5-20\) keV) \citep{nelson1995}.  In addition, \cite{doroshenko2012} found that the high-energy component is more energetic than expected from \cite{nelson1995}.  


The fundamental cyclotron energy can provide an estimate of the B-field using \(E_B = 11.6 B_{12} \) keV (ref).  In the CRSF model, the fundamental cyclotron energy is 27 keV.  In contrast, the fundamental cyclotron energy is roughly 51 keV.  Thus, the B-field values are \( \sim 2 \times 10^{12}\) G and \( \sim 4 \times 10^{12}\) G for the CRSF and cyclotron emission line models, respectively.  These values are similar to values reported for other accretion powered pulsars, which typically have a few \( \times 10^{12}\) G (See \cite{staubert2019,kim2023}).


As is shown in Figure~\ref{fig:avg_spec}, the four models differ above \( \sim 90\) keV.  The \texttt{CompTT+CompTT} model is unable to describe the high-energy data, which disfavors that model.  The three \texttt{comptb} models describe the data nearly equally well, with difficulties in each of their interpretations.  Therefore, we are not able to differentiate between them based on our spectra, though higher significance results above \(\sim 120 \) keV could exclude some models.

\subsection{Comparison with other sources}

Recent results of Be/X-ray binaries during low-luminosity states have found spectra comparable X Per.  These sources also exhibit significantly different luminosities to investigate how the spectra evolve.  In the case of GX \(304-1\), using \textit{Swift}/XRT and \textit{NuSTAR} data, \cite{tsygankov2019a} reported a comparable two-hump spectrum that they interpreted as likely due to either CSRF or cyclotron emission.  Additionally, results of GRO J\(1008-57\) with \textit{SRG}ART-XC and \textit{NuSTAR} at different luminosities show an evolution of the spectrum, with the second hump becoming prominent at the lowest luminosity, which is consistent with a CRSF line at \(80-90\) keV \citep{lutovinov2021}.  

1A \(0535+262\) provides a different picture as its luminosity decreases.  It shows a two-hump spectrum in its lowest luminosity like GRO J\(1008-57\) and GX \(304-1\), but in \textit{Swift}/XRT and \textit{NuSTAR} the source also has an additional spectral feature modelled as a Gaussian at \( \sim 48\) keV and a width of roughly 13 keV \citep{tsygankov2019b} that is not seen in the other sources.  The presence of a cyclotron line suggests that the high-energy hump in 1A \(0535+262\) is not a CRSF and that the features seen in the other sources are also not CSRFs.  As bulk Comptonization emission is unlikely to be efficient at the low accretion rates found in low-luminosity states since the required optical depths are not present \citep{mushtukov2015}, \cite{tsygankov2019a} excluded bulk Comptonization as an explanation for the high-energy hump in GX \(304-1\) during the source's low-luminosity state.  Thus, the authors interpreted the feature as a cyclotron emission line.  They propose a scenario without the cold atmosphere assumption of \cite{nelson1995}.  

\section{Conclusions}
In this work, we analyzed \textit{INTEGRAL} observations from \textrm{JEM-X} and \textrm{ISGRI} spanning roughly 15.5 years.  During this time, X Per displayed temporal variability in both instruments.  A search for spectral variability using periods with both \textrm{JEM-X} and \textrm{ISGRI} did not find any significant variability in the \(3-250\) keV energy range on a revolution timescale.  Also, an investigation of spectral variability at different count-rate levels (Figure~\ref{fig:flux_level}) did not find any significant variability. Consequently, we created an average spectrum using all the \textit{INTEGRAL} observations.  This \textit{INTEGRAL} spectrum was combined with a \textit{NuSTAR} spectrum to study the different spectral models.  

The origin of the hard X-ray emission in X Per has long been debated, with most of the discussion focused on the presence (or lack) of a CRSF line.  In order to address this question, we focused on physical models, instead of phenomenological ones. 
 The analysis found that the two-hump spectrum can be described well by \texttt{comptb+comptb}, \texttt{comptb*gabs}, and \texttt{comptb+gauss} models.  A fit with \texttt{CompTT+CompTT} is statistically acceptable, but significantly over predicts the high-energy flux above \(\sim 90\) keV.  

The three \texttt{comptb} models result in comparably good fits; thus, we are not able to determine if the high-energy hump is due to Comptonization, cyclotron absorption, or cyclotron emission.  However, the expected fluxes for the three models begin to diverge above approximately 120 keV  (See Figure~\ref{fig:avg_spec}).  Thus, future observations to increase the high-energy statistics may be able to differentiate between the models.  

Bulk Comptonization is unlikely to be efficient in such low-luminosity states as the accretion rate is low \citep{mushtukov2015}.  Thus, it is unlikely to be the origin of the high-energy emission in X Per. The observed width of the \texttt{gabs} feature is also significantly broader than expected for a CRSF interpretation, which disfavours a CRSF interpretation in the case of X Per.  Therefore, we suggest that the high-energy hump seen in X Per is likely due to a cyclotron emission line.  

Recent observations of other Be/X-ray binaries during low-luminosity states have found similar spectral shapes (GX 304\(-\)1, GRO J1008\(-\)57, and 1A 0535\(+\)262), indicating that the behavior seen in X Per is not unusual.  \cite{tsygankov2019b} results for 1A 0535\(+\)262 report a CRSF line on top of the high-energy hump, which suggests that the hump is not a cyclotron line, further supporting the interpretation of the high-energy feature in X Per and the other sources as being due to a cyclotron emission line.

\begin{acknowledgements}
      The authors thank the referee for their comments and input.  The authors thank the Italian Space Agency for the financial support under the “INTEGRAL ASI-INAF” agreement n◦ 2019-35-HH.0.  The research leading to these results has received funding from the European Union’s Horizon 2020 Programme under the AHEAD2020 project (grant agreement n. 871158).  This research has made use of data and/or software provided by the High Energy Astrophysics Science Archive Research Center (HEASARC), which is a service of the Astrophysics Science Division at NASA/GSFC.   This work has made use of data from the European Space Agency (ESA) mission {\it Gaia} (\url{https://www.cosmos.esa.int/gaia}), processed by the {\it Gaia} Data Processing and Analysis Consortium (DPAC, \url{https://www.cosmos.esa.int/web/gaia/dpac/consortium}). Funding for the DPAC has been provided by national institutions, in particular the institutions participating in the {\it Gaia} Multilateral Agreement.
\end{acknowledgements}

%
%

\end{document}